\documentclass[aps,pra,twocolumn,superscriptaddress,showpacs,amsmath,amssymb,flushbottom]{revtex4-1} 

\usepackage{graphicx}
\usepackage{amsfonts,amsmath,amssymb}
\usepackage[normalem]{ulem}

\newcommand{\bop}[1]{\hat{b}_{#1}} 
 
\newcommand{\bdop}[1]{\hat{b}^{\dagger}_{#1}} 
\newcommand{\nop}[1]{\hat{n}_{#1}}
\newcommand{\veck}{ \veck }

\newcommand{\bj}[2]{ \mathcal{J}_{#1}(#2)} 
\newcommand{\fw}{F_{\omega}} 
\newcommand{\w}{\omega} 
\newcommand{\ka}{\kappa} 
\newcommand{\tf}{t_{f}} 
\newcommand{\hef}{\hat{H}_{\textrm{eff}}} 
\newcommand{\hfl}{\mathfrak{H}(t)} 
\newcommand{\ham}{\hat{H}} 
\newcommand{\je}{J_{\textrm{eff}}} 
\newcommand{\aver}[1]{\leftangle #1\rightangle} 
\def \aver#1{{\langle#1\rangle}}

\def \ket#1{{|#1\rangle}}
\def \bra#1{{\langle #1|}}

\begin{document} 

\title{Slow quench dynamics of periodically driven quantum gases}  

\author{Dario Poletti}

\affiliation{D\'epartement de Physique Th\'eorique, Universit\'e de Gen\`eve, 24, quai E. Ansermet 1211 Gen\`eve 4, Suisse}  

\author{Corinna Kollath$^{1,}$} 

\affiliation{Centre de Physique Th\'eorique, Ecole Polytechnique, CNRS, 91128 Palaiseau, France}

\date{\today} 

\begin{abstract}
We study the evolution of bosons in a periodically driven optical lattice during a slow change of the driving amplitude. 
Both the regime of high frequency and low frequency driving are investigated. In the low frequency regime, resonant absorption of energy is observed. 
In the high frequency regime, the dynamics is compared to a system with an effective Hamiltonian in which the atoms are `dressed' by the driving field. 
This `dressing' can dramatically change the amplitude and sign of the effective tunneling.  
A particular focus of this study is the investigation of the time-scales necessary for the evolving quantum state to follow almost adiabatically to the ground-state of the effective many body system.  
\end{abstract}


\maketitle

\section{Introduction} 

 In experiments with cold atomic gases confined in optical lattices, many parameters can be tuned almost at will \cite{BlochZwerger2008}. This makes them a very promising set-up in order to study long standing questions as the origin of high temperature superconductivity, or the 
existence of spin liquids in frustrated quantum magnets.

Recently the use of high frequency periodical driving forces has been proposed to engineer complex effective many body Hamiltonians not reachable by other means. The effect of periodic forces to `dress' atoms is well known  from quantum optics where potentials for atoms can be designed by the application of light fields \cite{Foot2005,CohenTannoudjiGrynberg1998}. 
The dressing is not only efficient for non-interacting atoms. For weakly interacting Bose-Einstein condensates, an effective suppression of tunneling in double well 
structures \cite{KierigOberthaler2008} and also in one-dimensional optical lattice potentials \cite{LignierArimondo2007} as been observed. 
In the context of strongly interacting atom gases, interesting effects of high frequency driving forces are predicted as for example the change of sign of the tunneling amplitude in an optical lattice 
\cite{EckardtHolthaus2005, EckardtHolthaus2008} which can induce an effective negative temperature \cite{TsujiAoki2010}, controlled bound-pair transport \cite{KudoMonteiro2009}, the building of frustrated antiferromagnetism at experimentally feasible temperatures 
\cite{EckardtLewenstein2010} and the generation of effective staggered magnetic fields and hence unconventional superconductivity \cite{LimHemmerich2010}. 
An important step towards the design of such complex quantum phases by driving potentials in strongly interacting systems was the recent experimental realization of the transition from a superfluid to a Mott-insulator by 
accelerating the lattice \cite{ZenesiniArimondo2009}. This proof of principle experiment proposed in Ref.~\cite{EckardtHolthaus2005} shows the feasibility of the generation of effective models in atomic gases with 
strong interactions despite the occurrence of many-body resonances.  

However, important questions to address are not only the engineering of complex effective Hamiltonians, but also how well for example the ground state of the effective Hamiltonian could be reached by slowly switching on the driving potential and how much heating the driving potential induces. To answer these questions the almost perfect isolation of cold atom gases by which no thermalization with any surrounding environment takes place has to be taken into account. Due to this isolation the preparation of the atomic cloud is almost entropy conserving (and temperature is not conserved), if the changes of parameters are performed slowly enough. 
The question of whether it is possible to follow the ground-state of a strongly correlated state while slowly changing the parameters of the non-periodically driven system is still an unresolved puzzle \cite{Dziarmaga2010, PolkovnikovVengalattore2010}. 
In the context of bosonic atoms confined in an optical lattice, it has been addressed recently both theoretically \cite{CanoviSantoro2009, BernierKollath2010,NatuMueller2011} and experimentally \cite{HungChin2010,BakrGreiner2010}. 
The presence of many avoided and unavoided level crossings in the many-body spectrum makes an understanding very involved. 
For the crossing of a phase transition to an ordered phase the dynamics has been proposed to obey a universal mechanism, the so called Kibble-Zurek mechanism \cite{Kibble1976,Zurek1985}. 
This mechanism predicts an algebraic dependence of the defect generation on the ramp time. However, its validity in quantum systems is not yet established \cite{Dziarmaga2010,PolkovnikovVengalattore2010}. 

Even more involved is the question of adiabaticity in a periodically driven many body system \cite{HoneKohn1997}. 
The description of the periodic driving is often performed using the so-called Floquet formalism \cite{Shirley1965,GrifoniHanggi1998}. This formalism uses the fact that due to the periodicity in time, a Brouillon-zone-like structure occurs in the quasi-energy spectrum. 
Whereas in a non-driven system the energy values follow smoothly a change in a time-varying parameter, the Floquet quasi-energies are found to exhibit discontinuities in the thermodynamic limit during the change of the amplitude of the driving \cite{HoneKohn1997}. 
Therefore strictly speaking an adiabatic limit does not exist. 
However for many physical relevant situations of large system sizes the quasi-energies are expected to remain close to the values of the undriven system \cite{HoneKohn1997}.

Previous numerical studies in many body systems have investigated the behavior of a periodically driven system for small system sizes (approx. $12$ sites) \cite{EckardtHolthaus2005}. 
Different excitations depending on the driving frequency have been identified due to the crossing of (avoided) level crossings in the energy spectrum. 
In the regime of high frequency driving, a decent following of the ground state value has been pointed out. 
An improvement of the adiabatic following has been suggested by a shortening of the turn-on time, since in this way, at narrow crossings, Landau-Zener transitions could be suppressed \cite{EckardtHolthaus2005}. 

It is not clear how these findings in small systems carry over to larger system sizes in which finite size gaps decrease and dense energy bands develop. In our work we will investigate the response of a bosonic atom cloud of realistic size confined to an optical lattice potential to a periodic driving while slowly increasing the driving amplitude to a desired value. 
As a driving force we consider the tilting of the optical lattice potential which experimentally can be generated by the use of a detuning of the frequencies \cite{LignierArimondo2007} or a spatially oscillating mirror reflecting the lattice beams \cite{IvanovTino2008}. 
We investigate three different physical situations: (i) the tuning to a vanishing tunneling amplitude, (ii) the decrease of the tunneling amplitude to a final value without changing the sign and (iii) the inversion of the sign of 
the tunneling amplitude. This choice is motivated by different physical situations. Due to the direct proportionality between the force induced and the mass of the atoms, the driving could be used as an efficient method to manipulate the tunneling constant of only selected species in an optical lattice 
(i.e. the heavier ones). Thereby generating a vanishing effective tunneling (i) could be used to create a quenched disordered potential for the heavier species. 
Situation (ii) is a more general change of a parameter in a system, interesting, for example, in the study of a superfluid to Mott-insulator transition. 
Finally situation (iii), i.e. the inversion of the sign of the tunneling amplitude, is a very interesting tool to engineer more complex quantum phases as spin liquids in frustrated geometries \cite{EckardtLewenstein2010}. 
Our study of the inversion of sign in a simple one-component bosonic system will be the test-bed if this inversion is possible. In particular we investigate whether it is possible to produce a state which is close to the 
ground-state of the effective Hamiltonian without producing too many excitations when inverting the sign of the tunneling parameter.

\section{Setup} 
In experiments with cold atomic gases, potentials of different geometries can be engineered by the application of light or magnetic fields. Here we consider the application of two far-detuned counter-propagating 
laser beams to a one-dimensional tube of bosonic atoms. The atoms feel a potential $V(x,t)$ of the form: 
\begin{equation}
V(x,t)=V_0\sin^2\left( k_L (x+X(t) ) \right). 
\end{equation}
Here $k_L=2\pi/\lambda_L$ is the wave vector and $\lambda_L$ the wavelength of the laser, $t$ denotes the time. The potential strength $V_0$ can be tuned experimentally, since it is proportional to the intensity of the laser. 
Additionally to the standing wave usually created to form a steady periodic potential, $k_L X(t)$ is a time-dependent phase which can be adjusted by introducing a dephasing of the two counter-propagating lasers, e.g. by 
a moving mirror. We consider that the phase varies periodically in time, i.e.~$X(t)=A(t)\cos(\w t)$ (where $A(t)$ is the amplitude of the oscillations). 

Considering a sufficient height of the periodic potential, many physical properties of the ultracold bosonic atoms can be described within a one-band Bose-Hubbard Hamiltonian 
\cite{FisherFisher1989, JakschZoller1998,EckardtHolthaus2005}
\begin{eqnarray}
 &&\hat{H} =\hat{H}_\textrm{tun}+\hat{H}_{\textrm{int}}+ \fw(t) \sum_j a j\;\nop{j} \label{eq:Hamiltonian}\\
&&\textrm{where}\quad \hat{H}_\textrm{tun}= - J \sum_j (\bdop{j}\bop{j+1}+h.c.)\nonumber\\
&&\textrm{and}\quad \hat{H}_{\textrm{int}}= \frac U 2 \sum_j \nop{j}(\nop{j}-1). \nonumber
\end{eqnarray}

The operator $\bop{j}$ ($\bdop{j}$) annihilates (creates) a boson at site $j$. The operator $\hat{n}_j= \bdop{j}\bop{j}$ is the number operator on site $j$ and $a$ denotes the lattice spacing. 
The first term models the kinetic energy of the atoms with the 
tunneling amplitude $J$ and the second term the short range interaction between the atoms with the onsite interaction strength $U$. 
The third term, which describes a force periodically varying in time with frequency $\omega$, stems from the detuning of the counter-propagating laser beams.
This force is related to the phase oscillations of the detuned laser beams by 
\begin{eqnarray} 
&&\fw (t) = -m_0\frac{d^2X(t)}{dt^2} \\ 
    &&= m_0 \left[ \w^2 A(t)\cos(\w t) + 2\frac{dA}{dt} \w \sin(\w t) - \frac{d^2 A}{dt^2} \cos(\w t) \right]. \nonumber 
\end{eqnarray}
where $m_0$ is the mass of an atom. If the amplitude $A(t)$ is varying slowly compared to $\w$ the expression takes the simpler form \footnote{We checked numerically the results of the simplified against the full expression. 
Some deviations are found in particular for small driving frequencies. However they do not affect our results.} 
\begin{equation} 
\fw(t)= f(t)\cos(\w t) \quad \textrm{with} \quad f(t)=A(t) m_0 \w^2.    
\end{equation} 
We will investigate this model for realistic system sizes $L$ around $48$ to $64$ sites with open boundary conditions. We use the time-dependent density matrix renormalization group method (t-DMRG) 
\cite{Vidal2004,DaleyVidal2004,WhiteFeiguin2004,GobertSchuetz2004,Schollwoeck2005} keeping up to a few hundred states if necessary. 

Before we turn to the description of the driven Bose-Hubbard model, let us summarize some of the properties of the Bose-Hubbard model in the absence of the driving, i.e.~$F_{\w}=0$. 
For weak interaction strength ($U/J<(U/J)_c$) a superfluid state is present, in which atoms are delocalized over the system. 
In one-dimension, two main characteristics of this state are an asymptotic algebraic decay of the single-particle correlation functions with distance and a gapless excitation spectrum. 
In contrast at strong interactions ($U/J>(U/J)_c$) a Mott-insulating state occurs at commensurate filling. 
In this state the atoms are localized in space, i.e. the single-particle correlation functions decay exponentially with distance and particle number fluctuations are strongly suppressed. 
The Mott-insulating state is incompressible and has a gap to the first excitations. For large interaction strengths, the gap is of the order of the interaction amplitude $U$. 
At incommensurate filling the system exhibits a crossover between the superfluid and the so-called Tonks-Giradeau gas at large interaction strengths. 
Note that in this limit of strong but finite interaction strength, many low energy properties can be described by a gapless fermionic system. However, in dynamical situations higher energy bands can play an important role.  

The approximation of taking only a single-band Bose-Hubbard model (\ref{eq:Hamiltonian}) is only valid as long as excitations to the higher lying bands can be neglected. 
In particular an energy quantum $ \hbar \omega$ of the driving field must be far-resonant with the excitation energies to the next bands. 
In the harmonic approximation, the band gap between the lowest and first excited Bloch band of the lattice is given by $\w_{ho}\approx 2 E_R\sqrt{V_0/E_R}$. 
Here we will only consider driving frequencies $\w$  lower than $\w_{ho}$. 

\section{Floquet description and effective Hamiltonian} 
Neglecting the slow increase of the driving amplitude, the considered system falls into the wide class of periodic Hamiltonians $H(t)=H(t+T)$, where $T$ is the period and $\omega=2\pi/T$ the frequency. 
This periodicity leads to specific properties which can be discussed within the so-called Floquet formalism \cite{Shirley1965,GrifoniHanggi1998}. 
In the case of a fast driving (we will specify later on what we mean by fast) an effective time-independent Hamiltonian can give a very good description of many physical properties. 
In the considered case at a fixed time $t$ this effective Hamiltonian \cite{EckardtHolthaus2005} is given by
\begin{eqnarray} 
 \hef &=& - J \bj{0}{f(t) a/\hbar \w} \sum_{j}(\bdop{j}\bop{j+1}+h.c.) \nonumber \\ 
&& + \frac U 2 \sum_j \nop{j}(\nop{j}-1)  \label{eq:Ham_eff} 
\end{eqnarray} 
where $\mathcal{J}_{n}$ denotes the Bessel function of n-th order.  Comparing to the Hamiltonian (\ref{eq:Hamiltonian}) without driving an effective tunneling amplitude has been generated by the driving which we will denote by 
$J_{\textrm{eff}}(t)\equiv J \bj{0}{f( t) a/\hbar \w}$.

 We recapitulate in the following the derivation of this effective Hamiltonian, since it gives many useful insights into the behavior of the system also outside the range of validity of the effective model.

For a Hamiltonian periodic in time, a complete set of solutions to the Schr\"odinger equation $\left( \ham - i\hbar \partial_t \right)|\psi\rangle=0 $ is given by the states $|\psi\rangle=|u_n(t)\rangle e^{-iE_n t/\hbar}$ 
where $|u_n(t)\rangle$ are $T-$periodic solutions of the eigenvalue problem 
\begin{equation} 
\hfl|u_n(t)\rangle=E_n|u_n(t)\rangle.  \label{eq:Floquet}  
\end{equation} 
Here $\hfl$ is the Floquet Hamiltonian which is given by $\hfl=\ham - i\hbar \partial_t$, $E_n$ are different quasi-energies and $|u_n(t)\rangle$ are the so-called Floquet states. If  $|u_n(t)\rangle$ is a Floquet state, the 
state $|u_n(t),m\rangle=|u_n(t)\rangle e^{i m \w t } $ (with $m$ integer and $\w=2\pi/T$) is also a $T-$periodic solution of equation \eqref{eq:Floquet} with quasi-energy $E_n+m\hbar\w$. 
Hence, in analogy to spatially periodic systems, the quasi-energy spectrum has a repeating Brillouin-like structure with the width of the Brillouin zone being $\hbar \omega$.  

The aim is to find a time-independent effective Hamiltonian, $\hef$, with an energy spectrum approximating the quasi-energy band $E_n\in [-\hbar\w/2,\hbar\w/2]$ of the Floquet Hamiltonian $\hfl$ averaged over one period of time. 
In order to derive this effective Hamiltonian, one performs a transformation of the basis states $|u_n(t)\rangle$ to a new opportunely chosen basis. 
In general, the new basis can be built using a $T-$periodic Hermitian operator $\hat{F}(t)=\hat{F}^{\dagger}(t)=\hat{F}(t+T)$ by $|n(t),m\rangle=U_{F,m}(t)|u_n(t)\rangle$ with $U_{F,m}(t)=e^{-i\hat{F}(t)+i m \w t } $ 
and $m\in \mathbb{Z}$.
The matrix elements of the Floquet Hamiltonian time-averaged over one period become \cite{Hemmerich2010}:
\begin{eqnarray}
&&\aver{\bra{n(t),m}\hfl\ket{n'(t),m'}}_T= \nonumber\\
&&\delta_{m,m'}\bra{n}(m\hbar \w +\aver{\hfl^{(0,0)}}_T \ket{n'} 
\\
&& +(1-\delta_{m,m'}) \bra{n}\aver{e^{i(m'-m)\w t}\hfl^{(0,0)} }_T\ket{n'} \label{eq:Floquet_offdiag0}
\end{eqnarray}
Here we used the notation $\hfl^{(m,m')}\equiv U^{\dagger}_{F,m}(t) \hfl U_{F,m'}(t)$ and $\langle ... \rangle_T$ denotes an average over one period $T$. 
The energy spectrum of the  Hamiltonian $\hef= \langle \hfl^{(0,0)} \rangle_T$ well approximates the quasi-energy spectrum of the time-averaged Floquet Hamiltonian $\hfl$ if the coupling between different bands can 
be neglected. 

In the considered situation Eq.~(\ref{eq:Hamiltonian}), a good choice of the operator is $\hat{F}(t)=(f a/\hbar \omega) \sin(\omega t)\sum_j j n_j $. 
Starting from the basis formed by the Fock states $\ket{\{n_l\}}$ the new basis becomes $ \ket{\{n_l\},m}=\ket{\{n_l\}} e^{-i  (f a/\hbar\w) \sin(\w t) \sum_j j n_j +i m\w t}$. 
The matrix elements of the time-averaged Floquet Hamiltonian become
\begin{eqnarray}
&&\aver{\bra{\{n'_l\},m'}\hfl\ket{\{n_l\},m}}_T= \nonumber\\
&&\delta_{m',m}\bra{\{n'_l\}}m\hbar \w + \hat{H}_{\textrm{int}}+j_0\hat{H}_\textrm{tun} \ket{\{n_l\}}
\label{eq:Floquet_diag}
\\
&&+(1-\delta_{m',m})j_{s(m-m')} \bra{\{n'_l\}}\hat{H}_\textrm{tun}\ket{\{n_l\}}
\label{eq:Floquet_offdiag}
\end{eqnarray}
where $j_\alpha=\mathcal{J}_{\alpha}(f a/\hbar\w)$ and $s=\sum_j j(n_j-n'_j)$.   

We see that if the non-diagonal coupling vanishes, $\hbar \w \gg U$ or $J$, and only one zone $E_n\in [-\hbar\w/2,\hbar\w/2]$ is taken into account, we recover the effective Hamiltonian (\ref{eq:Ham_eff}). 
This means that this effective Hamiltonian describes well the properties of the system if no absorption event described by the off-diagonal part takes place. It is indeed important to understand, in experimentally interesting problems for one-dimensional gases, when this approximation is valid. 
We will discuss the non-diagonal term more in detail in the next section, since beside its importance for the validity of the effective Hamiltonian it also reveils a lot of information on the resonant absorption at low frequencies.


\section{Resonant versus effective behavior} 
In this section we discuss the behavior of the system subject to a driving with a wide range of frequencies. Motivated by recent experimental work \cite{LignierArimondo2007,ZenesiniArimondo2009}, 
in which the driving amplitude was increased linearly with time, we will --if not stated otherwise-- focus on a linear increase of the amplitude of the phase detuning $f(t)=f_t t$. 
Therefore we do not deal with an exact time-periodic Hamiltonian anymore. However, we found that if the change of the amplitude is very slow compared to the driving frequencies, 
the properties of periodic systems are still a reasonable approximation. 

\begin{figure}[!ht] 
\includegraphics[width=0.8\columnwidth ]{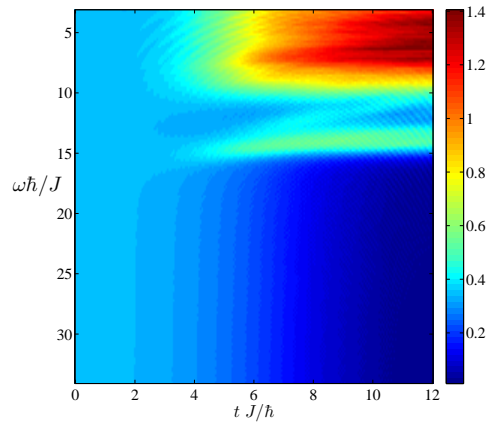} 
\caption{(color online) Time-evolution of the compressibility $\ka$ for different driving frequencies $\w$. The value of the strength of the driving force $f(t)$ is linearly increased in time, such that at time 
$t_f=12\hbar/J$, $\bj{0}{f_t t_f a /\hbar \w}=0 $. The `undressed' interaction strength is $U/J=4$ and system length $L=48$ with $N=48$ atoms. } \label{fig:CompVsTime}   
\end{figure}

A typical response of the system subject to the driving, is summarized in Fig.~\ref{fig:CompVsTime}. Note, that these are results of the exact time-evolution using the t-DMRG method.  
Fig.~\ref{fig:CompVsTime} shows how the compressibility $\ka_j=\langle \nop{j}^2 \rangle-\langle \nop{j} \rangle^2$ at $j=L/2$ evolves in time when the driving amplitude increases. In order to have a lighter notation, we will use $\ka\equiv\ka_{L/2}$. 
The final driving amplitude is chosen, such that the final value of the effective tunneling vanishes (case (i)). 
At the final time of the evolution the compressibility of the ground-state of the effective Hamiltonian (from now on we will refer to the ground-state of the effective Hamiltonian at the final driving strength as the target-state) is zero. 
Let us also note that the compressibility at uniform filling is related to the interaction energy of the system by $E_I = U/2 \sum_j \kappa_j+\textrm{const}$. 
Therefore the absorption of interaction energy is directly reflected in the compressibility. 

\begin{figure}[!ht] 
\includegraphics[width=\columnwidth ]{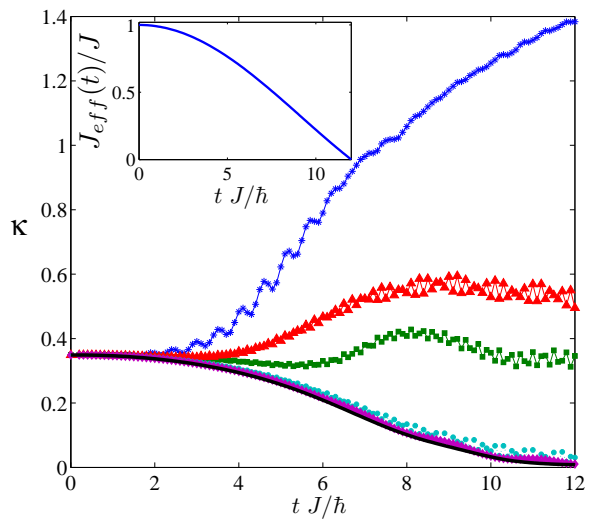} 
\caption{(color online) Time-evolution of the compressibility $\ka$ for chosen driving frequencies $\w$ ($\hbar\w=6J$, stars; $\hbar\w=12J$, squares; $\hbar \w=14J$, triangles; $\hbar\w=18J$, circles; $\hbar \w=30$, diamonds) 
and without driving following the decrease of the effective tunneling amplitude (continuous line). The inset shows the evolution of the effective tunneling amplitude. Same parameters as in Fig.~\ref{fig:CompVsTime}. } \label{fig:CompVsTime2}   
\end{figure}

For low driving frequencies a resonant absorption of energy, i.e.~a strong increase of  $\ka$, in certain frequency bands occurs (cf.~Fig.~\ref{fig:CompVsTime2} $\hbar\w/J=6,14$). 
In contrast for large frequencies a slow {\it decrease} of $\kappa$ is visible. This slow decrease appears to be almost independent of the driving frequency (cf.~Fig.~\ref{fig:CompVsTime2}, $\hbar\w/J=18$ and $\hbar\w/J=30$). 
This is the regime in which the behavior of the system can be described by the effective model (\ref{eq:Ham_eff}) taking at each time $t$ the effective tunneling amplitude $\je(t)$. 
The effective tunneling amplitude slowly decreases with time causing a decrease in the compressibility. This effective model compares reasonably well to the exact numerical calculations at high frequency (cf.~Fig.~\ref{fig:CompVsTime2}).   
Let us note that for some frequencies the time-evolution shows a mixed behavior between the two discussed cases. 
A crossover in time might occur between the decrease of the compressibility for small time, i.e.~the diagonal term dominates, and the absorption of energy for larger time, i.e.~the non-diagonal term dominates 
(cf.~Fig.~\ref{fig:CompVsTime2}, $\hbar\w/J=12$).

In the following we will discuss the occurrence of the resonances. Similar resonances have already been studied in different contexts \cite{IucciGiamarchi2006,KollathGiamarchi2006,ClarkJaksch2008}. In the considered model in the strong coupling limit resonant behavior has been investigated by linear response theory \cite{TokunoGiamarchi2011} and in the Floquet formalism \cite{EckardtHolthaus2008} investigating the off-diagonal term (Eq.~\eqref{eq:Floquet_offdiag}) in perturbation theory.  

A strong resonance occurs at $U \approx \hbar \omega$ corresponding in the large $U$ limit to a single particle-hole excitation. 
Weaker resonances describing multi-photon processes to the same single particle-hole excitation are expected at $m\hbar \omega \approx U$ ($m$ is an integer). 
The resonant behavior around $U \approx \hbar \omega$ is clearly seen in Fig.~\ref{fig:CompVsTime}. However, due to the chosen intermediate value of $U/J=4>(U/J)_c$ this resonance 
is very broad \footnote{In the strong coupling limit the width is expected to be $6J$.} and no higher photon resonance can be resolved. 
The coupling to the naively expected two particle-hole excitations at $m \hbar \omega \approx 2U$ vanishes up to 
second order in $J/U$ for systems at filling one \cite{KollathGiamarchi2006, EckardtHolthaus2008,TokunoGiamarchi2011}. In contrast, excitations at $3U$ are expected. 
In second order in $J/U$, only resonances at $m \hbar \omega \approx 3U$, where $m$ is an even integer,
 have non-vanishing matrix elements \footnote{The coupling to these resonances may be suppressed for certain values of the driving strength \cite{EckardtHolthaus2008}.}. 
In Fig.~\ref{fig:CompVsTime} a resonance at $2 \hbar \omega \approx 3U$ seems to be present leading to a substructure of the broad resonant region. 
Additionally to the predicted resonances, in Fig.~\ref{fig:CompVsTime}, a weaker resonance is seen which shifts from $\hbar \omega \approx 4U$ towards $\hbar \omega \approx 3U$. 
In the strong coupling limit this resonance corresponds to a process in which two particles from a singly occupied site hop to the same site of filling one between them (thus forming a triply occupied site). 
A shift of this resonance with the parameter $U/J$ has already been found in linear response calculations in the context of the modulation spectroscopy  \cite{ClarkJaksch2008}. 
Hence the resonance is related to the spectral properties of the undriven system. We further find that a weak branch separates from this weaker resonance  and joins the frequency band around $\hbar \w =2U$. 
We believe that this branch is due to subtle effects of the dressing of the original energy states by the strong driving. 
 Even though resonances at higher value of the frequency $\hbar \omega$ exist if higher orders in perturbation theory are considered, they seem to be very much suppressed for $\hbar \omega \gtrsim 5U$ at the shown parameters. 
In this regime the off-diagonal coupling terms (\ref{eq:Floquet_offdiag}) can be neglected and the effective description approximates the behavior of the system. 
The range of frequencies above which the effective regime is valid depends very much on the system parameters. 
For example at unit filling and low interaction strength $U/J\approx 1-2$ the frequency range above $\hbar\w=12 J$ is well described whereas at larger interaction strength, $\hbar\w>5U$ is a relatively safe approximation.

The behavior at resonance changes for different values of the filling \cite{KollathGiamarchi2006}. We found additional resonances and shifts in their position and weights at larger fillings. 
For example at filling $n=1.5$ and $n=2$, a strong resonance occurs in the compressibility at $\hbar \omega\approx 5U$. 
In the strong coupling limit this resonance can be associated with the tunneling of two atoms in a single occupied site to a doubly occupied site located between them. 
This forms a site with four atoms with its two neighboring sites non-occupied and has an energy difference to the ground-state of about $5U$. 
This process is highly suppressed for unit filling but becomes important for larger fillings since the probability to have a doubly occupied site strongly increases. 
In the experiments it will be important to avoid these additional resonances either by detuning the driving frequency or by assuring that the density in the trapping potential does not exceed a certain filling. 

Let us note that the resonant energy absorption destroys very quickly the coherence between the atoms which can be seen by a rapid decrease of the single-particle correlations (not shown).

\section{Slow quench dynamics in the high frequency regime} 
We now focus on the regime of high frequency driving. We aim to quantify the quality of the preparation of a target-state. 
The situation we consider is a slow linear increase of the driving amplitude $f(t)=f_t t $ over a finite ramp time $t_f$. 
This preparation by almost adiabatic increase is intriguing since in infinite systems the Floquet states and their quasi-energies have been shown to be discontinuous functions of the driving amplitude almost everywhere \cite{HoneKohn1997}. 
 
Previous numerical studies focused on small system sizes of at most 12 sites \cite{EckardtHolthaus2005,EckardtHolthaus2008}. 
In these small systems, finite size gaps occur and could lead to a very different behavior of the dynamics. 
We study the dependence of the proximity of the final state to the target-state on the ramp time $t_f$ in large, and experimentally relevant, systems.  
To distinguish the effect of the driving force from the adiabatic change in the effective tunneling amplitude, we compare the driven situation to the response of the system to a slow change of the tunneling amplitude without 
the presence of the driving (at each time the tunneling amplitude is chosen to correspond to the effective tunneling amplitude which would have been induced by the driving). 
The later situation will be called `undriven' in the following.   
Three physically different situations are distinguished (i) the preparation of a state with the vanishing effective tunneling ($\je(\tf)=0$), (ii) a state corrsponding to a reduced but finite effective tunneling ($\je(\tf)=J/2$),
 and (iii) a state in which the sign of the tunneling amplitude has been reversed ($\je(\tf)\approx -0.4J$).

\subsection{Slow decrease of the effective hopping when just in the Mott region} 

\begin{figure}[!ht] 
\includegraphics[width=\columnwidth ]{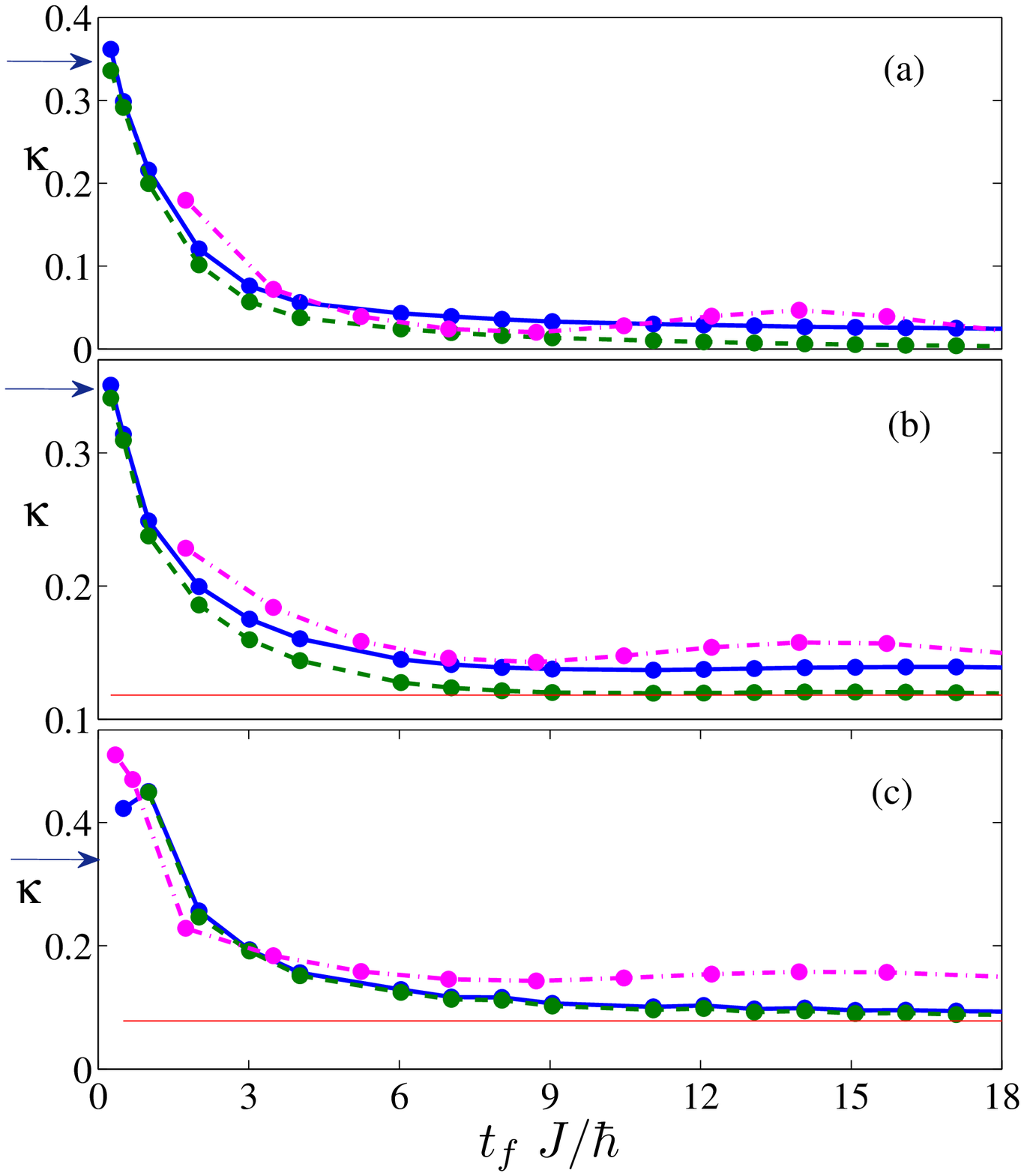} 
\caption{(color online) Compressibility $\ka$ versus the final time of the quench $\tf$ for different driving amplitudes $f_t$ such that (a) $\je(\tf)=0$ , (b) $\je(\tf)=0.5 J$, and (c) $\je(\tf)=-0.4J$. 
Different curves are $\hbar \w/J=18$ (dash-dotted line) and $\hbar\w/J=25$ (continuous line), undriven (dashed line). The symbols denote actual numerical data. 
The horizontal lines represent the values of the compressibility of the target-states and arrows indicate the value of the compressibility in the initial state. Parameters are $U/J=4$, $N=L=48$.} \label{fig:CompVsTFin}   
\end{figure} 

The initial state we consider here is the ground state at the interaction strength $U/J=4$ at the filling of one particle per site. 
This state lies just above the phase transition in the Mott-insulating phase (the transition, for a homogeneous system at unit filling, is at $U/J\approx 3.38$ \cite{KuehnerMonien2000,ZakrzewskiDelande2008}).   
In Fig.~\ref{fig:CompVsTFin} the final value of the compressibility $\kappa$ versus the final time of the effective quench is shown for the situations (i)-(iii). 
The behavior of different driving frequencies $\w$ is compared to the expected behavior from the undriven case.   
In an experiment the reduction of the tunneling amplitude can be directly performed by the increase of the optical lattice strength. However, the change of sign cannot be induced by the simple modulation of the lattice height. 

In the undriven system, for short ramp times $t_f$ the value of the compressibility shows a rapid decrease from the value of the initial state with the ramp time \footnote{A slight rise is found for very fast ramps}. 
For long ramp times $t_f$ the value of the compressibility reached in the undriven system is in all situations close to that of the target-state. 
In the driven systems the compressibility shows roughly a similar decrease. However, different features have to be commented.  
For case (i) and (ii) this decrease of the compressibility agrees well with the undriven system for small ramp times $t_f$, but deviates considerably for larger ramp times $t_f$. For a very large frequency $\hbar \w=25J$ an almost monotonic decrease is found for large ramp times, whereas $\hbar \w=18J$ shows oscillations. 

Therefore it is clearly harder to evolve adiabatically towards the expected ground state value in the presence of a driving force rather than in the undriven case. 
We attribute this finding to the many level crossings or avoided level crossings which occur in the quasi-energies in the Floquet picture. 
If these quasi-energy level crossings are too wide, the state cannot follow adiabatically and excitations are generated. 
For small systems it was proposed that a faster ramp time could, in some situations, help to stay in a certain state \cite{EckardtHolthaus2005}. 
Also for large systems \cite{HoneKohn1997} it is expected that short, i.e.~finite, ramp time would perform better than completely adiabatic ramp.   
In our system we do not observe a clear signature of this phenomenon for the examined ramp times. In general we observe an improvement of the quality of the state obtained (neglecting slow oscillations over the ramp time).  

Surprisingly for case (iii) the evolution of the driven system follows closely the evolution of the undriven system. 
This means that in this case (iii) the main deviations already stem from the internal structure of the energy spectrum of the undriven system. 

\begin{figure}[!ht] 
\includegraphics[width=\columnwidth ]{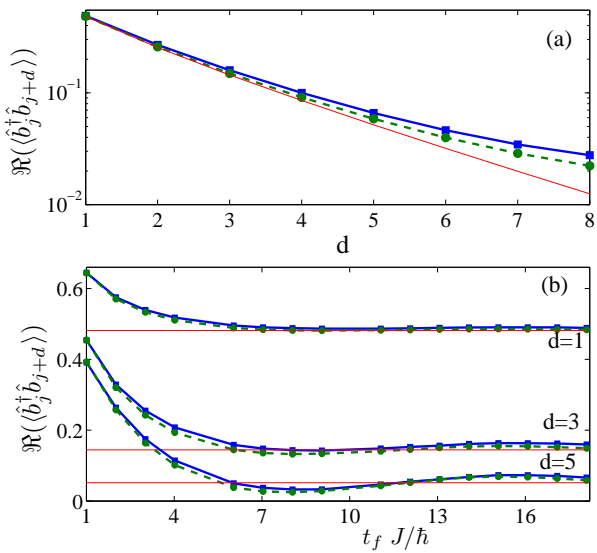} 
\caption{(color online)  (a) Single particle correlations between different sites versus their distance in the presence of the driving amplitude of frequency $\hbar\w/J=25$ (continuous line with squares) and for the undriven case (dashed line with circles). 
The target value is shown for comparison (continuous line). 
(b) Single-particle correlations ($d=1,3,5$) versus the final time of the evolution.  
Parameters:  $U/J=4$, $N=L=48$, $j=25$ and $f_t$ such that $\je(\tf)=0.5J $. Results are taken at exact periods of the driving.} \label{fig:CorrVsTFin}   
\end{figure}

Additionally to the local compressibility, we investigate the response of the single particle correlations $\langle \bdop{j}\bop{j+d} \rangle$ taken at the exact periods of the driving. For the ground-state of the superfluid (Mott-insulator) these have a characteristic asymptotic algebraic (exponential) decay with the distance $d$. In Fig.~\ref{fig:CorrVsTFin} (a) we show the decay of the single particle correlations for a fixed ramp time $t_f=12\hbar/J$. The target-state exhibits the exponential decay typical for correlations in the Mott-insulating regime. 
For small distances $d$ the correlations decay for both the driven and undriven cases to a value close to the value of the target state, whereas longer range correlations are harder to reach. In Fig.~\ref{fig:CorrVsTFin}(b) the dependence of the final value of the single particle 
correlations on the ramp time is shown for different distances $d$. Whereas the nearest-neighbor correlation ($d=1$) shows a decrease towards the value expected for the final ground state, the correlations for distances $d=3$ and $d=5$ show 
an oscillating behavior around the target-state value. Longer ramp times would be necessary to converge also these correlations. 

\begin{figure}[!ht] 
\includegraphics[width=\columnwidth ]{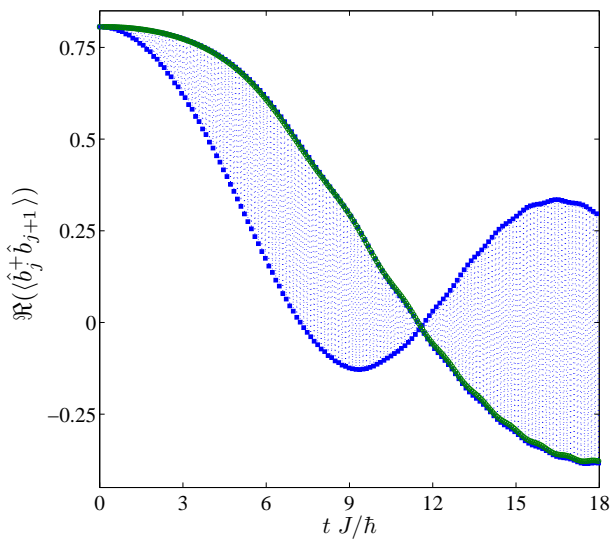} 
\caption{(color online) Real part of the nearest-neighbor correlation ($\Re(\langle \bdop{j}\bop{j+1} \rangle)$)  versus time for the driven case (dotted line) and undriven case (continuous line). 
Squares mark the values at each quarter of a period. Parameters: $U/J=4$, $\hbar\w/J=25$, $N=L=48$ and $f_t$ such that $\je(\tf)=-0.4 J$.  } \label{fig:KVstimeN}   
\end{figure}

In contrast to the compressibility, nearest-neighbor correlations (which are proportional to the kinetic energy) in the driven system show very strong oscillations in time with a frequency $\w_{osc}=2\w$. 
This is exemplified in Fig.~\ref{fig:KVstimeN} in which the nearest neighbor correlation is plotted versus time for the case (iii), i.e.~the case in which an effective change of the sign of $J$ occurs. 
The initial value of the correlation is very high, since the initial state is close to the superfluid phase. The correlation starts to decrease monotonically for times corresponding to the period of the 
oscillations $\w_{osc}$. However, with the increase of the driving amplitude strong oscillations build up. In fact the gas is accelerated periodically in opposite directions during one period of the driving potential.
Only close to the point where the effective tunneling vanishes these oscillations become small in amplitude. The value at the exact period becomes the minimum of the oscillations for longer times. 
Due to the fast oscillations the characteristics of the state is very different if a measure is not done at the exact periods of the driving. 
While the eigenenergies of the effective Hamiltonian $\hef$ do match closely the quasi-eigenenergies of the Floquet Hamiltonian $\hfl$, a lot of kinetic energy is absorbed and released during one period of oscillations. Hence in an experiment a very precise timing between the driving and the measurement must be assured.

\begin{figure}[!ht] 
\includegraphics[width=\columnwidth ]{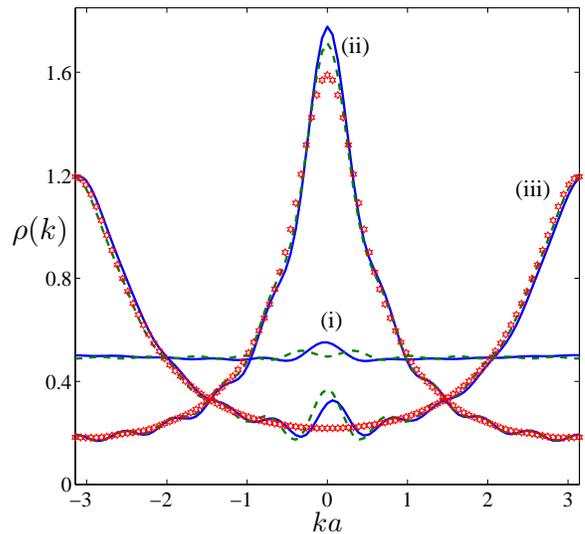} 
\caption{(color online) Momentum distribution $\rho(k)$ versus momentum for $f_t$ such that (i) $\je(\tf)=0 $, (ii) $\je(\tf)=0.5J $, (iii) $\je(\tf)\approx -0.4J $. 
The continuous line corresponds to the driven case, the dashed line to the undriven case and the stars to the value of the target-state. 
Parameters: $U/J=4$, $\hbar\w/J=25$, $N=L=48$, $t_f\approx 17\hbar/J$. } \label{fig:MomVstime}   
\end{figure}

In experiments one possible measurement technique is the time-of-flight imaging. For long time-of-flights, the signal of this measurement is related to the momentum distribution of the atoms which we show in Fig.~\ref{fig:MomVstime}. 
The momentum distribution is very different in the final target-states of the three considered situations \cite{EckardtHolthaus2007,ZenesiniArimondo2009}. 
If $\je$ is reduced to half its value, the initially sharp peak at $k=0$ broadens. At $\je=0$ the distribution becomes flat and if $\je$ changes sign a peak at $k=\pm\pi /a$ appears. 
The final distribution for a slow ramp time is shown in Fig.~\ref{fig:MomVstime}. 
This characteristic behavior can be clearly identified in the momentum distribution of the driven and undriven evolution (taken at the exact period). 
However, reminiscent features of the original distribution remain visible as a small peak at $k=0$ both in the driven and undriven situations.
 The details of these features are slightly distinct for the driven and undriven system. 
In agreement with the deviation found for longer range correlations, the remaining peak at $k=0$ is an experimentally detectable sign that the system has not yet reached the target-state, but is in an excited state. 

\subsection{Variation of the initial state}

\begin{figure}[!ht] 
\includegraphics[width=\columnwidth ]{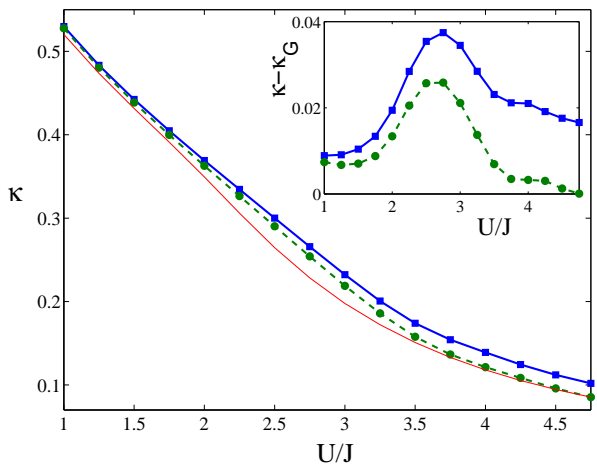} 
\caption{(color online) Compressibility $\ka$ versus the strength of the interaction $U$.  
Squares represents the values of the driven case, circles the undriven case and the continuous line the value of the target-state. Inset: Deviations of the compressibility $\kappa$ from that of the target state $\kappa_G$. 
Parameters:  $L=N=64$, $t_f\approx 12\hbar/J$, $\hbar\w/J=25$ and $f_t$ such that $\je(t_f)=0.5J $. } \label{fig:KVsU}    
\end{figure}

Until now we have focused on the case in which the initial state ($U/J=4$) lies just beyond the critical value inside the Mott-insulator regime. 
However, since the superfluid and Mott-insulating state typically have very distinct dynamics, it is interesting to study different quench scenarios: from superfluid to superfluid, from superfluid to Mott-insulator and from 
Mott-insulator to Mott-insulator. In Fig.~\ref{fig:KVsU} the final compressibility after a quench from $\je=J$ to $\je=J/2$ is compared for the undriven, driven and target-state starting at different initial values of $U/J$. 
The final compressibility for the undriven case is always closer to the one of the target-state than the one of the driven case. 
Moreover for intermediate initial values of $U/J$ ($1.7\lesssim U/J\lesssim 3.4$), that lead to the crossing of the phase transition, the deviations of both the driven and the undriven case from the target state are largest 
(see inset of Fig.~\ref{fig:KVsU} \footnote{For all these points, the error bars are smaller than the symbols used}). 
We attribute these deviations to the change of the nature of low lying energy levels around the phase transition. 
In the undriven case the deviations almost vanish for a quench inside the Mott-insulating phase and seem to be slightly more pronounced in the superfluid phase. 
The driven case shows larger deviations in the Mott-insulating region than for a quench within the superfluid phase.

\subsection{Maintaining the prepared state} 

\begin{figure}[!ht] 
\includegraphics[width=\columnwidth ]{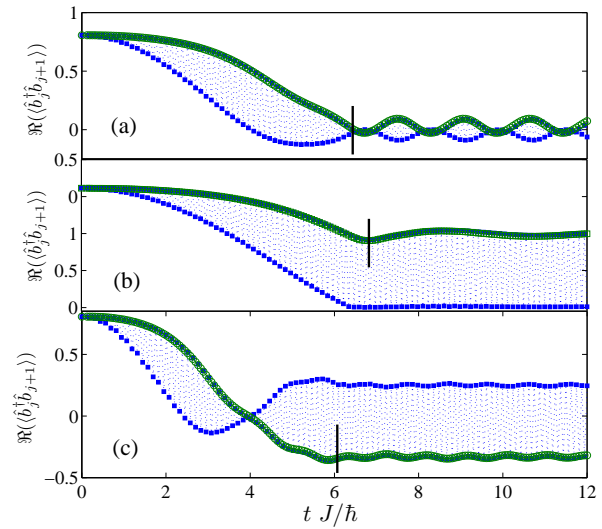} 
\caption{(color online) Real part of the nearest-neighbor correlation ($\Re(\langle \bdop{j}\bop{j+1} \rangle)$)  versus time for the driven case with $\hbar\w/J=25$ (dotted line, squares mark the value at a quarter of a period) and undriven case (continuous line). 
Parameters: $U/J=4$,  $N=L=48$, $j=25$ and $f_t$ such that (a) $\je(t)=0 $,  (b) $\je(t)=0.5J $,  (c) $\je(t)\approx-0.4J$ for all $t>t_f\approx 6\hbar/J$. $t_f$ is 
marked, in each case, with a vertical line. } \label{fig:KineticVsTFinHold}   
\end{figure} 

One of the purposes of using high frequency drivings is the preparation of complex quantum states. 
We have seen that for a slow increase of the driving amplitude a state close to the ground state of an effective Hamiltonian can be prepared in the sense that many properties of the driven state (measured exactly at the period of the driving) are close to the properties of the target-state. 
However, the question arises if this prepared many body state, which is continuously subject to a high frequency driving, is stable enough over a time necessary for experiments. 
In Fig.~\ref{fig:KineticVsTFinHold} we show the time evolution of the nearest-neighbor correlation for the cases in which the target-state is that of an effective Hamiltonian with $\je=0$ in (a), $\je=J/2$ in (b) and $\je\approx -0.4J$ in (c). In all these three cases we linearly increase the amplitude of the driving to the desidered value in an intermediate ramp time $t_f\approx 6\hbar/J$. After this time the amplitude of the driving force is kept constant. 
We find that the value of the nearest-neighbor correlations, during this time of constant amplitude of the driving, oscillates around a quasi-steady value (when measuring exactly at the period of the driving). 
The quasi-steady state seems to be reasonably stable on experimentally relevant time-scales and no energy absorption originating from the driving is evident. The amplitude of the oscillations depends on the performed quench. 
In particular for the quench towards an effectively vanishing tunneling amplitude the oscillations seem to be more pronounced than for the other two cases.
These oscillations stem from the non-differentiable form of the protocol for the driving amplitude and are well known from the finite time Landau Zener transition. 
Note that the derivative of the Bessel function at its first zero is large, such that the abrupt change of the oscillations amplitude from linearly increasing to constant in time leads to the largest oscillations in case (i). 
The oscillations might be minimized using smoother protocols for the change of the driving amplitude.

\section{Conclusions} 

We have investigated for experimentally realistic sizes the dynamics of bosonic atoms subject to a periodic driving force. 
At small driving frequencies, the resonant behavior can be employed to obtain information on the energy spectrum of the underlying quantum phase. 
At high driving frequencies an effectively undriven Hamiltonian is expected to describe the properties of the system at the exact driving frequencies. 
Increasing slowly the driving amplitude can be used to prepare the corresponding ground states of the complex effective Hamiltonians. 
We have found a wide range of frequencies, depending on the properties of the underlying state, in which an effective Hamiltonian can be well simulated by the high-frequency driving. 
The adiabatic following of local quantities while increasing the driving amplitude, however, has been found to be less efficient than in the undriven case and across a phase transition. 
Longer range quantities as the single particle correlation functions are more involved to reach than local quantities leading to experimentally detectable signals in the interference patterns. 
Finally we have found that once reached the desired effective Hamiltonian, the created state can be maintained over a certain time without sizable heating.  
In summary, the ground-state of an effective Hamiltonian can be prepared to a good approximation using a high driving force. 
This is very promising for the engineering of more complex states by the application of a driving force. 
However, the quality of the prepared state is very sensitive to the preparation process and in particular the time employed to increase the amplitude of the oscillations. 
This will be important when trying to produce states with particularly low entropy as spin liquid states in frustrated systems.\\  

{\it Note added} -- During the final stages of the preparation of this manuscript, frustrated systems have been realized experimentally (as show on the arXiv \cite{StruckSengstock2011}). 
In the light of these frustrated systems the question of whether the ground state properties of a driven system can be exploited becomes even more urgent. \\ 

{\it Acknowledgments} -- We would like to thank T. Giamarchi, O. Morsch and A. Tokuno for fruitful discussions. We acknowledge support provided by the `Triangle de la Physique', ANR (FAMOUS), DARPA-OLE, the Swiss National Foundation. 
Additionally, financial support for the computer cluster on which the calculations were performed has been provided by the `Fondation Ernst et Lucie Schmidheiny'.



\end{document}